\documentclass[twocolumn,a4paper,amsfronts,showpacs]{revtex4}
\usepackage{amsmath,amsthm}
\usepackage{graphics}
\usepackage{epsfig}
\begin{document}
\title {Dynamical systems theory for nonlinear evolution equations}
\author{Amitava Choudhuri$^1$, B Talukdar$^1$}
\email{binoy123@bsnl.in}
\author{Umapada Das$^2$}
\affiliation{$^1$ Department of Physics, Visva-Bharati, Santiniketan 731235, India}
\affiliation{$^2$ Department of Physics, Abhedananda Mahavidyalaya, Sainthia 731234, India}
\begin{abstract}
We observe that the  fully nonlinear evolution equations of Rosenau and Hymann, often abbreviated as $K(n,\,m)$ equations, can be reduced to  Hamiltonian form only on a zero-energy hypersurface belonging to some potential function associated with the equations. We treat the resulting Hamiltonian equations by the dynamical systems theory and present a phase-space analysis of their stable points. The results of our study demonstrate that the equations can, in general,  support both compacton and soliton solutions. For the $K(2,\,2)$ and $K(3,\,3)$ cases one type of solutions can be obtained from the other by continuously varying a parameter of the equations. This is not true for the $K(3,\,2)$ equation for which the parameter can take only negative values. The $K(2,\,3)$ equation does not have any stable point and, in the language of mechanics, represents a particle moving with constant acceleration.
\end{abstract}
\maketitle
\vskip 0.5 cm
\noindent{\bf 1. Introduction}
\vskip 0.5 cm
\noindent The theories of dynamical systems via phase-space analysis provide a very useful tool to extract information about any physical system modelled by linear or nonlinear differential equations. Here the long term behaviour of the system is investigated in terms of the stability of fixed points of the differential equations. In particular, one examines how the stability or instability is affected as the parameter of the model or differential equation is varied. This viewpoint was  followed by Caffey {[}1{]} to identify the closed form solution of a number of quasilinear differential equations with special attention to phase space trajectories that connect a critical point to itself or other critical points.\par A nonlinear evolution equation is often called quasilinear when the dispersive term of it is linear. In addition to quasilinear equations there are fully nonlinear evolution (FNE) equations characterized by nonlinear dispersive terms. The FNE equations play a role in the studies of pattern formation and wave breaking phenomena {[}2{]}. As a straightforward generalization of the quasilinear KdV equation Rosenau and Hyman {[}3{]} introduced a class of FNE equations given by
$$u_t-A(u^n)_x+(u^m)_{xxx}=0\,\,.\eqno(1)$$
Equations in $(1)$ are often called $K(n,m)$ equations with $n>0\,\,,\,\,1<m\leq 3$. Here $A$ is a parameter of the model. The FNE equations in $(1)$ are non integrable and support compacton solutions. The compacton represents a travelling solitary wave solution with compact support. The object of the present work is to provide a phase-plane analysis of $(1)$ and thus illustrate the type of insight one can gain from the study. \par The quasilinear equations can easily be treated by the dynamical systems theory. For example, the change of variable 
$$u(x,\,t)=\phi(x-vt)=\phi(\xi)\eqno(2)$$
followed by an integration over $\xi$ converts the quasilinear KdV equation
$$u_t+\alpha uu_x+\gamma u_{xxx}=0\eqno(3)$$
with parameters $\alpha$ and $\gamma$, to an ordinary nonlinear differential equation
$$\frac{d^2\phi}{d\xi^2}=\frac{v}{\gamma}\phi-\frac{\alpha}{2\gamma}\phi^2\eqno(4)$$
such that $\phi$ plays the role of space coordinate and $\xi$, that of a time coordinate. Thus for stability analysis, $(4)$ can be viewed as an equation of anharmonic oscillator in the variables $\phi$ and $\xi$.
\vskip 0.5 cm
\noindent{\bf 2. Analogue of $(4)$ for FNE equations}
\vskip 0.5 cm
\noindent For the FNE equations in $(1)$, a similar change of variable from $u$ to $\phi$ and subsequent integration over $\xi$  lead to ordinary differential equation
$$\frac{d^2\phi}{d\xi^2}=\frac{v}{m}\phi^{2-m}+\frac{A}{m}\phi^{n-m+1}-\frac{m-1}{\phi}(\frac{d\phi}{d\xi})^2\,\,.\eqno(5)$$
The last term in the right side of $(5)$ does not permit one to regard these equations as Hamiltonian systems and give rise to an awkward analytical constraint for the application of dynamical systems theory.  This problem can, however, be resolved by considering evolution of $\phi$ in a zero-energy hypersurface in the phase space belonging to some potential function.\par The potential representation for the analysis of travelling wave solutions of nonlinear dispersive evolution equations was introduced by Eichmann, Ludu and Draayer {[}4{]}. This representation is defined by
$$(\frac{d\phi}{d\xi})^2=-{\cal F}(\phi)\,\,.\eqno(6)$$
The left side of $(6)$ was identified with the non-relativistic kinetic energy and right side with the negative value of a potential energy ${\cal F}(\phi)$ because $\xi$ and $\phi$ in $(2)$ could be regarded as time and space coordinate respectively. Here $\frac{d\phi}{d\xi}$ is the velocity of a particle moving along the $\phi$-axis. Clearly, for the potential representation in $(6)$, the evolution of $\phi$ or $u$ proceeds on the zero-energy hypersurface in the phase space belonging to ${\cal F}(\phi)$. The function ${\cal F}(\phi)$ for the $K(n,\,m)$ equation in $(1)$ is given by {[}4{]}
$${\cal F}(\phi)=-\frac{2A}{m(m+n)}\phi^{n-m+2}-\frac{2v}{m(m+1)}\phi^{3-m}\,\,.\eqno(7)$$
Using $(7)$ in $(5)$ we get
$$\frac{d^2\phi}{d\xi^2}=\frac{v}{m}\left(1-\frac{2(m-1)}{(m+1)}\right)\phi^{2-m}$$$$+\frac{A}{m}\left(1-\frac{2(m-1)}{(m+n)}\right)\phi^{n-m+1}\,\,.\eqno(8)$$
Equation $(8)$ can be written in the newtonian form
$$\frac{d^2\phi}{d\xi^2}=-\frac{dV(\phi)}{d\phi}\eqno(9)$$ with the potential
{\small$$V(\phi)=-\frac{v}{m(3-m)}\left(1-\frac{2(m-1)}{(m+1)}\right)\phi^{3-m}$$$$-\frac{A}{m(n-m+2)}\left(1-\frac{2(m-1)}{(m+n)}\right)\phi^{n-m+2}\,\,.\eqno(10)$$}
The second-order differential equations in $(8)$ are equivalent to two first-order equations given by 
$$\frac{d\phi}{d\xi}=\psi={\cal P}(\phi,\,\psi),\,\,{\rm(say)}\eqno(11a)$$
and
$$\frac{d\psi}{d\xi}=\frac{v}{m}\left(1-\frac{2(m-1)}{(m+1)}\right)\phi^{2-m}$$$$+\frac{A}{m}\left(1-\frac{2(m-1)}{(m+n)}\right)\phi^{n-m+1}={\cal Q}(\phi,\,\psi),\,\,{\rm(say)}\,\,.\eqno(11b)$$
Here $\psi$ stands for the velocity of the particle at a point $\phi(\xi)$ on the hypersurface defined by $(6)$. It is rather straightforward to show that $(11a)$ and $(11b)$ form a Hamiltonian system with the Hamiltonian
$$H(\phi,\,\psi)={1\over 2} \psi^2+V(\phi)\,\,.\eqno(12)$$
Clearly, the canonical equations
$$\frac{d\phi}{d\xi}=\frac{\partial H}{\partial\psi}\,\,\,\,\,\,\,\,\,\,{\rm and}\,\,\,\,\,\,\,\,\,\,\,\frac{d\psi}{d\xi}=-\frac{\partial H}{\partial\phi}\eqno(13)$$
lead to $(8)$. Formally, the critical  points of $(11)$ are given by
$$(\phi,\,\,\psi)\equiv\left((-\frac{\alpha}{\beta})^{\frac{1}{n-1}},\,0\right)\eqno(14)$$
where
$$\alpha=\frac{v}{m}\left(1-\frac{2(m-1)}{(m+1)}\right)$$ and
$$\beta=\frac{A}{m}\left(1-\frac{2(m-1)}{(m+n)}\right)\,\,.\eqno(15)$$
\vskip 0.5 cm
\noindent{\bf 3. Phase-space analysis of $K(n,\,m)$ equations}
\vskip 0.5 cm
\noindent The travelling wave solutions of $(1)$ are usually studied by restricting the values of both integers, $n$ and $m$ to $2$ and $3$. We shall, therefore, present the phase-space analysis of the $K(n,\,m)$ equations given by
$$K(2,\,2)\,:\,\,\,\,\frac{d^2\phi}{d\xi^2}=\frac{v}{6}+\frac{A}{4}\phi=-\frac{d}{d\phi}\left(-\frac{v}{6}\phi-\frac{A}{8}\phi^2\right)\,\,,\eqno(16a)$$
$$K(2,\,3)\,:\,\,\,\,\frac{d^2\phi}{d\xi^2}=\frac{A}{15}=-\frac{d}{d\phi}\left(-\frac{A}{15}\phi\right)\,\,,\eqno(16b)$$
$$K(3,\,2)\,:\,\,\,\,\frac{d^2\phi}{d\xi^2}=\frac{v}{6}+\frac{3A}{10}\phi^2=-\frac{d}{d\phi}\left(-\frac{v}{6}\phi-\frac{A}{10}\phi^3\right)\eqno(16c)$$
and
$$K(3,\,3)\,:\,\,\,\,\frac{d^2\phi}{d\xi^2}=\frac{A}{9}\phi=-\frac{d}{d\phi}\left(-\frac{A}{18}\phi^2\right)\eqno(16d)$$
with a view to study the stability of their critical points. The equation of motion $(16d)$ can be viewed as that of an inverted harmonic oscillator while $(16a)$ represents a similar oscillator with shifted origin. In the context of Bose-Einstein condensates the potentials in $(16a)$ and $(16d)$ are often referred to as expulsive potential {[}5{]}. Equation $(16b)$ gives the motion of a particle in a linear potential. For small $\phi$, the potential in $(16c)$ is similar to that of $(16b)$. For large $\phi$, however, the potential is highly nonlinear. \par From $(11)$ it is easy to see that, as with linear harmonic oscillator, $(0,\,0)$ is also the critical point of $(16d)$. It is well known that the critical point of linear oscillator is a center and is neutrally stable. We shall now study the stability of critical point of $(16d)$. Since all $K(n,\,m)$ equations can be written in the form of $(11)$ we make use of the eigenvalues of the Jacobian matrix
$$M=\left(\begin{array}{cc}a&b\\c&d\end{array}\right)\,\,,\,\,\,\,\,\,\,\,\,\,\,\,\,{\rm det}M\neq 0\,\,\eqno(17)$$
for the classification of their critical points. In $(17)$ 
$$a=\frac{\partial{\cal P}}{\partial \phi}|_{(\phi^*,\,\psi^*)}\,\,\,,\,\,\,b=\frac{\partial{\cal P}}{\partial \psi}|_{(\phi^*,\,\psi^*)}$$
and
$$c=\frac{\partial{\cal Q}}{\partial \phi}|_{(\phi^*,\,\psi^*)}\,\,\,,\,\,\,d=\frac{\partial{\cal Q}}{\partial \psi}|_{(\phi^*,\,\psi^*)}$$
where $(\phi^*,\,\psi^*)$  stands for the critical point. For $(16d)$ the eigenvalue equation is given by
$$\left|\begin{array}{cc}-\lambda&1\\\frac{A}{9}&-\lambda\end{array}\right|=0\eqno(18)$$
which gives $\lambda_{1,2}=\pm\frac{\sqrt{A}}{3}$ such that the equilibrium point $(0,\,0)$ is a saddle for $A>0$.
\begin{figure}
\includegraphics[width=6.8cm]{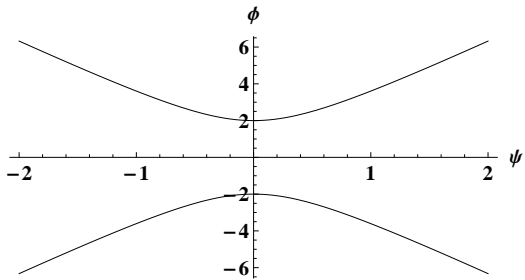}
\caption{Phase diagram for the $K(3,\,3)$ equation}
\end{figure}
Since $\psi$ is the time derivative of $\phi$, the plot of $\phi$ versus $\psi$ will define the phase-plane for the equation.  The integral curve for the $K(3,\,3)$ equation as obtained from $(11)$ is a hyperbola
$$\phi^2-{9\over A}\psi^2=c,\,\,\,{\rm a\,\, constant.}\eqno(19)$$
Figure $1$ gives the phase plane for $A=1$ and a particular choice of $c$, say $4$. The hyperbolic trajectories reach the equilibrium point along two directions only and in all other directions the trajectories diverge from it. Thus we can say that, in general, the trajectories diverge from the critical point making it unstable. 
\begin{figure}
\includegraphics[width=6.8cm]{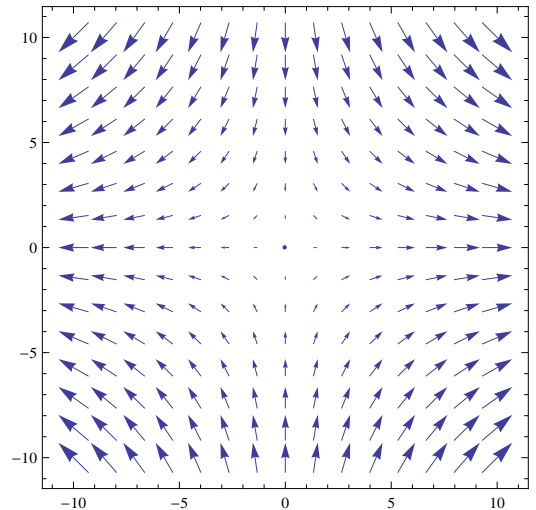}
\caption{Vector flow for the integral curve in $(19)$}
\end{figure}

In Figure $2$ we plot the vector (or tangent) field for the $K(3,\,3)$ equation resulting from the integral curve $(19)$. Clearly, at all points the tangent vectors diverge from the critical point $(0,\,0)$ and reconfirm that it is an unstable equilibrium of the system. For $A<0$, the critical point becomes a centre. A local bifurcation occurs when a parameter change causes the stability of the equilibrium point to change. Thus the system modelled by the $K(3,\,3)$ equation is expected to exhibit bifurcation as the values of $A$ are changed from negative to positive.\par For the critical point at the origin and $A>0$ we write $(\phi,\,\psi)=(\Phi,\,\Psi)$. Specializing $(11)$ to the $K(3,\,3)$ case we linearize the system about $(0,\,0)$ and write the matrix equation
$$\frac{d}{d\xi}\left(\begin{array}{c}\Phi\\\Psi\end{array}\right)=\left(\begin{array}{cc}0&1\\{A\over 9}&0\end{array}\right)\left(\begin{array}{c}\Phi\\\Psi\end{array}\right)\,\,.\eqno(20)$$
Since the eigenvalues of the square matrix in $(20)$ are $\pm\frac{\sqrt{A}}{3}$ we can write the general solution of the linear equation as
$$\left(\begin{array}{c}\Phi\\\Psi\end{array}\right)=a_1\left(\begin{array}{c}1\\\frac{\sqrt{A}}{3}\end{array}\right)e^{\frac{\sqrt{A}}{3}\xi}+a_2\left(\begin{array}{c}1\\-\frac{\sqrt{A}}{3}\end{array}\right)e^{-\frac{\sqrt{A}}{3}\xi}\eqno(21)$$
where $a_1$ and $a_2$ are constants of integration determined by the initial values $\Phi(0)$ and $\Psi(0)$. The $K(2,\,2)$ equation in $(16a)$ is similar to that in $(16d)$ except that we have an inverted oscillator with shifted origin. It is quite straightforward to show that our analysis for the $K(3,\,3)$ equation also applies to $(16a)$.\par As opposed to $K(2,\,2)$ and $K(3,\,3)$ equations, there is no simple relationship between the $K(2,\,3)$ and $K(3,\,2)$ equations. Curiously enough, reduction of the $K(2,\,3)$  equation to two first-order equations does not define a critical or stable point of the system. Rather, $(16b)$ gives the motion of a particle with constant acceleration. On the other hand, we can easily write such first-order equations for the $K(3,\,2)$ equation giving the critical points $(\pm{i\over3} \sqrt{5\over A},\,0)$ for $v=1$. An important implication of this result is that $A$ should always be less than zero in $(16c)$. The choice $A=-1$ leads to the critical points  $(\pm{\sqrt{5}\over 3},\,0)$. The eigenvalues of the Jacobian matrix of this problem at $({\sqrt{5}\over 3},\,0)$ are purely imaginary while those at $(-{\sqrt{5}\over 3},\,0)$ are real and have opposite signs such that the former critical point is a centre while the later one is a saddle. The integral curve as obtained from the first-order equations corresponding to $(16c)$ is given by 
$$\phi^3-{5\over 3} \phi+ 5 \psi^2=0\,\,.\eqno(22)$$
The solutions $\phi=f(\psi)$ of this equation will give the phase trajectory. Since $(22)$ is cubic in $\phi$, we can write its solutions as
$$\phi_1=\alpha=\frac{2^{1/3} 5^{2/3}}{3 \left(-27 \psi^2+\sqrt{-20+729\psi^4}\right)^{1/3}}+ $$$$\frac{1}{3} \left(\frac{5}{2}\right)^{1/3} \left(-27 \psi^2+\sqrt{-20+729\psi^4}\right)^{1/3}\,\,,\eqno(23a)$$
$$\phi_2=\beta=-\frac{\left(\frac{5}{2}\right)^{2/3} \left(1+i \sqrt{3}\right)}{3 \left(-27 \psi^2+\sqrt{-20+729\psi^4}\right)^{1/3}}- $$$$\frac{1}{6} \left(\frac{5}{2}\right)^{1/3}\left(1-i \sqrt{3}\right) \left(-27\psi^2+\sqrt{-20+729 \psi^4}\right)^{1/3}\eqno(23b)$$
and
$$\phi_3=\gamma=-\frac{\left(\frac{5}{2}\right)^{2/3} \left(1-i \sqrt{3}\right)}{3 \left(-27 \psi^2+\sqrt{-20+729 \psi^4}\right)^{1/3}}-$$$$\frac{1}{6} \left(\frac{5}{2}\right)^{1/3}\left(1+i \sqrt{3}\right)\left(-27\psi^2+\sqrt{-20+729\psi^4}\right)^{1/3}\,\,.\eqno(23c)$$
\begin{figure}
\includegraphics[width=6.8cm]{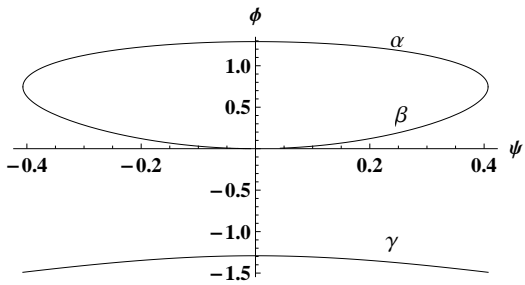}
\caption{Phase diagram for the $K(3,\,2)$ equation}
\end{figure}
It can be seen that for $-0.409682\leq\psi\leq0.409682$, all $\phi_i$'s are real else one of the $\phi_i$'s is real and other two are complex. For $|\psi|\leq0.409682$ we present, in figure $3$, the phase trajectories defined by $(23)$. The trajectories of $\alpha$ and $\beta$ form a closed curve and clearly show that the system has a centre at  $({\sqrt{5}\over 3},\,0)$, while $\beta$ and $\gamma$ curves are similar to those of figure $1$ such that the critical point  $(-{\sqrt{5}\over 3},\,0)$ is a saddle.
\begin{figure}
\includegraphics[width=6.8cm]{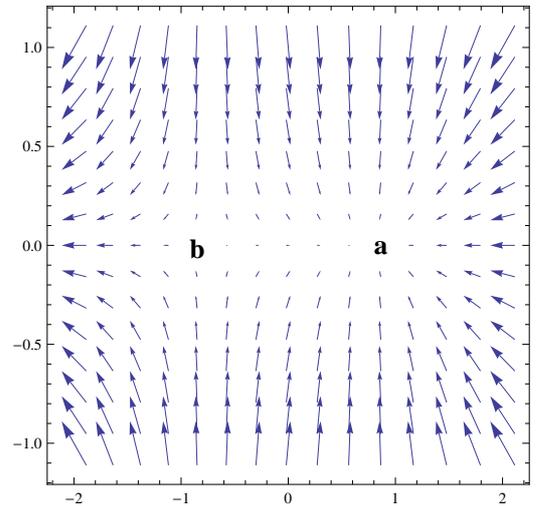}
\caption{Vector flow for the integral curve in $(22)$ }
\end{figure}
Figure $4$ portrays the vector field corresponding the integral curve $(22)$. Looking closely into this figure we see that the tangent vectors converge towards the critical point $({\sqrt{5}\over 3},\,0)$ labeled by ${\bf a}$ and diverge away from $(-{\sqrt{5}\over 3},\,0)$ labeled by ${\bf b}$ since they represent the stable and unstable equilibria of the system.\par Linearization of the system of equations implied by $(16c)$ about the centre $({\sqrt{5}\over 3},\,0)$ we get
$$\phi=a_1e^{i{({1\over5})^{1\over4}}\xi}+a_2e^{-i{({1\over5})^{1\over4}}\xi}\eqno(24)$$
which represents a compacton for $a_1=a_2=a\,\,\,\,{\rm (say)}$. On the other hand, similar linearization of the system about the saddle point $(-{\sqrt{5}\over 3},\,0)$ leads to a soliton solution.
\vskip 0.5 cm
\noindent{\bf 4. Conclusion}
\vskip 0.5 cm
The nonintegrable $K(n,\,m)$ equations of Rosenau and Hyman have been extensively studied in the literature with a view to construct their solutions which appear in the form of compacton, peakon, cuspon etc.. In this work we did not try to present any new method to solve these equations. On the contrary, we made use of the phase-plane analysis of dynamical systems theory to study the nature of the solutions and their interconnection. Thus one may reasonably ask whether our study could provide any added realism for the solutions of the equations. Our answer to this query is fairly straightforward. We worked with $K(n,\,m)$ equations characterized by a parameter $A$ that can continuously tune the effects of nonlinearity. We found that unlike evolution equations with linear dispersive terms the FNE equations do not form Hamiltonian systems. However, the later reduce to the Hamiltonian form on the zero-energy potential hypersurface. On this surface the $K(2,\,2)$ and $K(3,\,3)$ equations have two critical/equilibrium points, namely, a centre and a saddle, the presence of which lead to compacton and soliton solutions respectively. As the parameter $A$ is continuously varied from positive to negative values the compacton solution goes over to the soliton solution. This represents a signature of a local bifurcation. For the $K(3,\,2)$ equation the variation of $A$ is constrained to negative values only. But as with the other two equations the $K(3,\,2)$ equation is also characterized by two critical points leading once again to compacton and soliton solutions. But one cannot be reached from the other by varying $A$. The $K(2,\,3)$ equation describes motion of a particle with constant acceleration. We conclude by noting that these features of the solutions did remain undiscovered in any method for solving them, however efficient it might be. \\
\\
{\bf Acknowledgements} \\
\\
This work is supported by the University Grants Commission, Government of India, through grant No. F.32-39/2006(SR).
\vskip 0.5 cm
\noindent{\bf References}
\vskip 0.5 cm
\noindent[1] Coffey M W 1992 {\it SIAM J. Appl. Math.} {\bf 52} 929\\

\noindent[2] Li Y A, Olver P J and Rosenau P 1999, Non-analytic solutions of nonlinear wave models, in Nonlinear Theory of Generalized Functions, Vienna 1997, {\it Chapman $\&$ Hall/CRC Res. Notes Math.} {\bf 401}  Chapman $\&$ Hall/CRC, Boca Raton, FL, 129-145\\

\noindent[3] Rosenau P and Hyman J M 1993 {\it Phys. Rev. Lett.} {\bf 70} 564\\

\noindent[4] Eichmann U A, Ludu A and Draayer J P 2002 {\it J. Phys. A: Math. Gen} {\bf 35} 6075\\

\noindent[5] Carr L D and Castin Y 2002 {\it Phys. Rev. A} {\bf 66} 063602\\

\end{document}